\documentclass[10pt]{article}
\usepackage[utf8]{inputenc}
\usepackage{geometry}
\usepackage{cite}

\usepackage{nameref,hyperref}

\usepackage{changepage}

\usepackage[aboveskip=1pt,labelfont=bf,labelsep=period,singlelinecheck=off]{caption}
\usepackage{graphicx}

\begin{document}
\vspace*{0.35in}

\begin{flushleft}
{\Large
\textbf\newline{Frequency responses of the K-Rb-$^{21}$Ne co-magnetometer}
}
\newline
\\
Yao Chen
\\
\bigskip
\bf School of Instrumentation Science and Opto-electronics Engineering, Beihang University, Beijing, China
\\
\bigskip
yao.chen.cn@outlook.com

\end{flushleft}

\section*{Abstract}
The frequency responses of the K-Rb-$^{21}$Ne co-magnetometer to magnetic field and exotic spin dependent forces are experimentally studied and simulated in this paper. Both the relationship between the output amplitude, the phase shift and frequencies are studied. The responses of magnetic field are experimentally investigated. Due to a lack of input methods, others are numerically simulated.

\section*{Introduction}
Atomic co-magnetometers use two spin ensembles occupying the same volume to suppress their sensitivity to magnetic field noise and maintain the sensitivity to rotations\cite{kornack2005nuclear}, anomalous spin coupling with space anisotropy\cite{smiciklas2011new} anomalous spin forces\cite{vasilakis2009limits}
etc. In a K-Rb-$^{21}$Ne co-magnetometer, the alkali spin ensemble couples with the noble gas spin ensemble through spin exchange
interaction and alkali atoms are in the spin exchange relaxation free (SERF) regime\cite{kominis2003subfemtotesla} to sensitively detect the direction of the $^{21}$Ne nuclear spin.\\
As a sensor, the co-magnetometer's dynamical properties are very important. There are various literatures focus on the dynamics of the co-magnetometer to external magnetic field\cite{kornack2002dynamics,fang2016dynamics,fang2016low,kornack2005nuclear}. This study focus on the frequency response of the co-magnetometer to exotic spin dependent forces. As the coupling of the electron spins and the nuclear spins in the co-magnetometer to external exotic spin dependent forces are different. We independently investigate the effect of spin force on the electron spins and nuclear spins. The full Bloch equations are given and the responses of the co-magnetometer to spin forces are numerically solved. Results show that under relatively high frequencies, the electron spins and the nuclear spins should decouple due to their different dynamical properties. Under low frequencies, the electron spins and the nuclear spins should strongly couple to each other. 

\section*{Theory}
This simulation based on the full Bloch equations of the co-magnetometer. There are various literature describes the full Bloch equations for the K-Rb-$^{21}$Ne co-magnetometer\cite{chen2016spin,kornack2002dynamics}. Here we just use them to simulate the responses of the co-magnetometer to external fields. In this paper, we are interested in the co-magnetometer's responses to external spin dependent force. There are some argument that the spin dependent force could be treated as an effective magnetic field\cite{ji2016searching}. Different from the traditional magnetic field, the coupling of the forces to the electron spins and nuclear spins in the co-magnetometer are independent. So we should study separately the responses of the forces on electron spins and nuclear spins. For the frequency response, a sinusoidal signal will be applied to the Bloch equations to get the output amplitude and the relative phase shift between the input and output signal. \\ 
The choices of parameters in the co-magnetometer's Bloch equation for simulation is critical. In a K-Rb-$^{21}$Ne co-magnetometer, typically several atm $^{21}$Ne gases are utilized and a hybrid pumping technique is utilized to polarize $^{21}$Ne spins to approximately 20$\%$.We chose the parameters in this simulation based on the conditions in these references\cite{fang2016dynamics,fang2016low,chen2016spin}. The dynamics of the co-magnetometer are close related to the spin relaxation time of the electron spins and nuclear spins. Moreover, the electron spins and the nuclear spins will also coupled together under certain conditions. It is hard to give a analysis solution and that is the reason why we do some numerical simulations.\\
\begin{figure}[ht] 
\includegraphics[width=165mm]{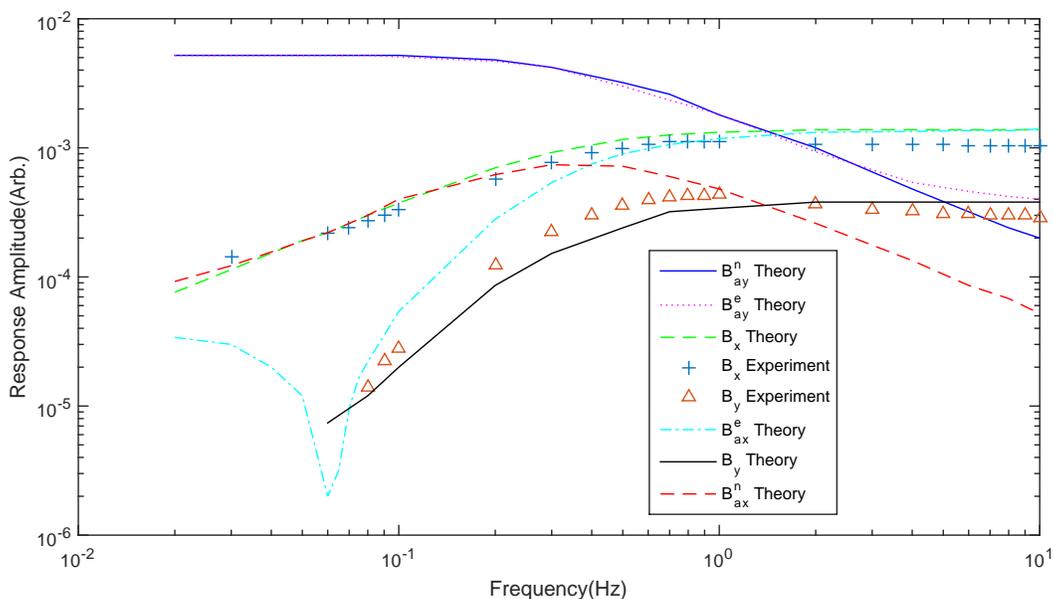}
\caption{\textbf{The output amplitude of the co-magnetometer under different input of magnetic fields under different frequencies.}}
\label{fig1} 
\end{figure}
\begin{figure}[ht]
\includegraphics[width=165mm]{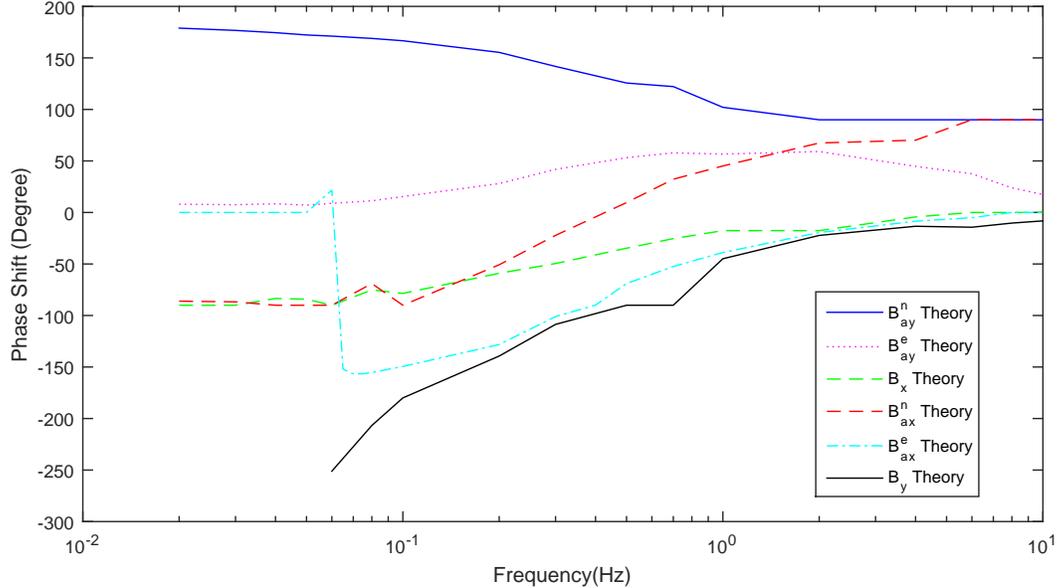}
\caption{\textbf{The relative phase shift between the co-magnetometer output and the input magnetic fields under different frequencies.}}
\label{fig2} 
\end{figure}
\section*{Experimental and Simulation Results}
Different frequencies of magnetic fields could be applied to the co-magnetometer and thus the frequency responses could be derived. However there is no method to apply an exotic spin dependent forces to the co-magnetometer at different frequencies. So only the simulation results are given. The experimental conditions in this paper is similar to this references\cite{fang2016dynamics}. We applied a sinusoidal magnetic field to the co-magnetometer. The amplitude of the magnetic field is 0.08nT. This magnetic field is much smaller than the equivalent magnetic field line width. Thus the co-magnetometer works in the linear area. Both the x and y direction magnetic fields are studied. As the co-magnetometer probes the x polarization of the alkali metal spins, we directly give the amplitude of the x polarization sinusoidal output signal. We also give the phase shift of the output polarization signal to the input magnetic fields. Figure \ref{fig1} shows the experimental results of the x and y magnetic fields response. We also fit the experimental results to the simulation results. They fit well with each other. This result gives a very good evidence that our simulation modal is right and the parameters in the modal are reasonable. The frequency response of the co-magnetometer to magnetic fields at low frequencies are not sensitive. At higher frequencies, the response is larger. This is a fact that the nuclear spins and the electron spin ensembles will coupled together at low frequencies to automatically compensating the input magnetic field. Figure \ref{fig2} shows the phase shift of the output signal to the input x and y magnetic field signal.\\
The responses of the co-magnetometer to exotic spin dependent forces are also studied in this paper. In the co-magnetometer,both the alkali metal electron spins and the noble gas nuclear spins are coupled to external exotic spin dependent forces. The co-magnetometer probe the x direction electron spins to sense weak field. The coupling between the electron spins and the exotic forces could be directly detected. For the nuclear spins, they coupled to the external exotic spin forces. At the same time, they will produce a magnetic field which could be sensed by the electron spins through spin exchange interaction. So the coupling between the nuclear spins and the exotic forces could only be indirectly detected. We define the coupling between the nuclear spins and the exotic spin dependent forces to be like an effective magnetic field in the x and y direction to be $B^n_{ax}$ and $B^n_{ay}$(a here stands for abnormal. These fields are abnormal fields.). Similarly we define $B^e_{ax}$ and $B^e_{ay}$ to be the effective magnetic field produced by the exotic forces which are coupled to the electron spins. We assume a sinusoidal exotic forces input with frequencies range from 0.05Hz to 10Hz for the simulation. Figure \ref{fig1} also shows the simulation of the co-magnetometer to the exotic forces. We study the responses of the electron spins and nuclear spins independently. We chose the amplitude input to be 0.08nT for the electron spins and the nuclear spins. The vertical axis represents the polarization of the electron spin in the x direction. Figure \ref{fig2} shows the phase shift of the output signal relative to the input fields. 
\section*{Discussion}
The electron spins and the nuclear spins both response to the magnetic field and the co-magnetometer is not sensitive to low frequency magnetic fields. This is the self compensating effect in the co-magnetometer. At higher frequencies, the electron spins and the nuclear spins decoupled and the electron spins will response to the magnetic field and the nuclear spins will gradually lost sensitivity to external magnetic field due to small dynamical range. The exotic spin dependent field are different for the electron spins and nuclear spins. We simulate them independently in this paper. The most sensitive term at 1-10Hz frequency range for the exotic field are the x component of the $B^e_{ax}$. We could use this to measure the exotic spin dependent force at higher frequencies. The benefits is that at higher frequencies noises are usually low.


\end{document}